# Improved ACD-based financial trade durations prediction leveraging LSTM networks and Attention Mechanism

Yong Shi[1, 2, 3], Wei Dai[1, 2, 3], Wen Long[1, 2, 3, *], Bo Li[1, 2, 3]

1 *School of Economics and Management, University of Chinese Academy of Sciences, No. 80 of Zhongguancun East Street, Haidian District, Beijing, 100190, P.R.China*
2 *Research Center on Fictitious Economy and Data Science, Chinese Academy of Sciences, No. 80 of Zhongguancun East Street, Haidian District, Beijing, 100190, P.R.China*
3 *Key Laboratory of Big Data Mining and Knowledge Management, Chinese Academy of Sciences , No. 80 of Zhongguancun East Street, Haidian District, Beijing, 100190, P.R.China*
\* Correspondence: longwen@ucas.ac.cn

**Abstract:** The liquidity risk factor of security market plays an important role in the formulation of trading strategies. A more liquid stock market means that the securities can be bought or sold more easily.  As a sound indicator of market liquidity, the transaction duration is the focus of this study. We concentrate on estimating the probability density function $p(\Delta t_{i+1}|G_i)$ where $\Delta t_{i+1}$ represents the duration of the (i+1)-*th* transaction, $G_i$ represents the historical information at the time when the (i+1)-*th* transaction occurs. In this paper, we propose a new ultra-high-frequency (UHF) duration modelling framework by utilizing long short-term memory (LSTM) networks to extend the conditional mean equation of classic autoregressive conditional duration (ACD) model while retaining the probabilistic inference ability. And then the attention mechanism is leveraged to unveil the internal mechanism of the constructed model. In order to minimize the impact of manual parameter tuning, we adopt fixed hyperparameters during the training process. The experiments applied to a large-scale dataset prove the superiority of the proposed hybrid models. In the input sequence, the temporal positions which are more important for predicting the next duration can be efficiently highlighted via the added attention mechanism layer.

**Keywords:** Duration Prediction, Deep Learning, LSTM, ACD, Hybrid Model

# 1. Introduction

Market liquidity refers to the degree to which an asset can be bought and sold easily for a fair price [1]. In other words, market liquidity can be regarded as the speed at which transactions can be concluded while maintaining a basically stable price [1]. Therefore, market liquidity risk is one of the most common factors considered by security investors especially by high frequency traders in building a trading strategy.

With the rapid development of computer storage technology, transaction by transaction financial trading data is accessible to researchers. Let $t_i$ stand for the time at which the $i$-th trade occurs, so that the duration between the (i+1)-*th* and i-*th* trade is $\Delta t_{i+1} = t_{i+1} - t_i$, which can directly measure the transaction speed of financial trading. The autoregressive conditional duration (ACD) model proposed by Engle and Russell has been the primary framework used for analyzing trading durations of ultra-high frequency (UHF) data, which are irregularly time-spaced and convey meaningful information [2]. In ACD models, the transaction duration is decomposed into the multiplicative product of two components, the conditional (expected) duration and the unexpected duration. The expected component is the portion of transaction duration that is linearly conditional on past durations, whereas the unexpected duration is the fraction of duration beyond that which could be predicted from past durations, and is usually characterized by an exponential distribution.

Based on the work of Engle and Russell [2], many works tried to improve the ability of capturing the relation between the conditional duration and the lagged durations. For example, the logarithmic version of ACD model was provided in [3], the threshold autoregressive conditional duration model was proposed in [4], the asymmetric autoregressive conditional duration model was put forward in [5], the smooth transition ACD model and the time-varying ACD model were introduced in [6]. There are also many other works focusing on choosing a suitable distribution to characterize the unexpected duration. The distributions which have been applied to the ACD models includes the generalized Gamma distribution in [7], generalized *F* distribution in [8], the mixture of two exponential distributions in [9], the regime-switching Pareto distribution in [10], the mixture of an exponential and a generalized beta of type2 (GB2) distribution in [11], etc. Like many other statistical models, the ACD family models require strong assumptions which are difficult to satisfy in realistic situations [12].

In recent years, machine learning methods have been widely applied to image identification and natural language processing problems. Compared with traditional statistical models, machine learning methods have looser model assumptions and better generalization ability. The artificial neural network (ANN), inspired by the biological neural network, is one of the most widely used machine learning methods. According to Universal Approximation Theorem [13], feedforward neural networks can approximate a Borel measurable function to any desired degree of accuracy if sufficiently many hidden units with arbitrary squashing functions are provided. Recurrent neural networks (RNNs) are a family of specially designed artificial neural network networks capable of extracting temporal information via the cycle architecture [14]. As the development in optimization techniques and computation hardware, RNNs have been widely used in many different domains recently [15]. To solve the vanishing gradient/exploding problem of simple RNNs, Hochreiter S. proposed the LSTM neural networks which can help us to utilize a longer sequence of historical information [16]. Although having the merit of strong fitting ability, LSTMs cannot provide probabilistic output compared with ACD family models.

Inspired by the work from Kristjanpoller and Minutolo [17], we propose a new architecture called LSTM-ACD to predict the UHF transaction durations by combing the ANN networks and ACD framework. We take a fully data driven approach to extend the mean equation of classic ACD models while retaining the probabilistic inference ability. In addition, attention layer is added into our model to make a visualization of the proposed network and to improve the interpretability. The proposed architecture is applied to real-world stock duration datasets. The result shows that the proposed model produces more accurate estimation and prediction, outperforming the classic ACD model on mean squared error and quantile estimation.

The remains of this paper is organized as follows: Section 2 introduces the methodology detailedly while Section 3 contains the experiment design and the corresponding results in this study. Section 4 concludes this paper and points out the possible direction of future research.

## 2. Methodology

In Section 2, the ACD framework is integrated with LSTM networks to propose a new

LSTM-ACD model for predicting the trading durations of UHF data. This section is organized as follows. Section 2.1 introduces the classic ACD model. Section 2.2 describes the proposed LSTM-ACD architecture in detail. In section 2.3, attention mechanism layer is utilized to unveil the internal mechanism of the proposed model.

## 2.1. ACD model

A classic ACD model assumes that the durations are conditionally exponentially distributed with a mean that follows an ARMA process [2]. As shown in formula(1), the duration $\Delta t_i$ between the i-*th* and (i-1)-*th* trade is the multiplicative product of $\mu_i$ and $\varepsilon_i$, which represents expected and unexpected portion of the transaction duration respectively. In the conditional mean equation, $\mu_i$ is linearly depends on the lagged durations and the lagged terms of itself. $p, q$ in formula (2) represents the lagged order.

$$\Delta t_i = \mu_i \varepsilon_i \tag{1}$$

$$\mu_i = \omega + \sum_{j=1}^{p} \alpha_j \Delta t_{i-j} + \sum_{j=1}^{q} \beta_j \mu_{i-j} \tag{2}$$

A major limitation of classic ACD model is the assumption that the variables in the conditional mean equation behave in strict stationarity and linearity, but the duration sequences are usually in a non-linear or non-stationary state. Hence, this paper intuitively extends the linear conditional mean equation to nonlinear case by LSTM networks due to the strong fitting ability of deep learning techniques.

## 2.2. The proposed Attention-LSTM-ACD model

**2.2.1 LSTM-ACD model**

It has been generally known that the LSTM cell is able to store information over longer time range compared with simple RNNs. As depicted in Figure 1, the information flow propagating across time steps is controlled by three LSTM gates: the forget gate, the input gate and the output gate.

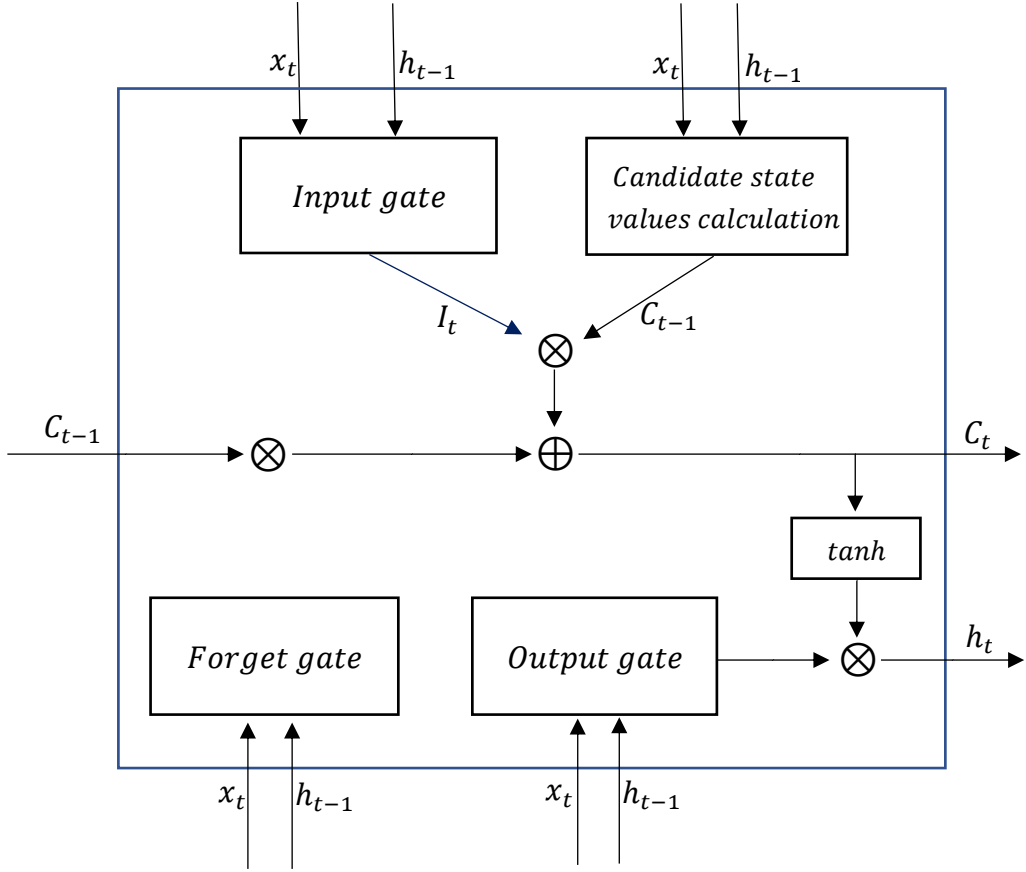

**Fig 1.** Structure of LSTM cell

Assuming that $W_f$, $W_i$, $W_o$, $W_c$ represent the LSTM weight matrices, $b_f$, $b_i$, $b_c$, $b_o$ represents the bias vectors. The input vector, output vector and cell state vector at time $t$ are denoted as $x_t$, $h_t$, $C_t$ respectively [18]. The operating process of a LSTM cell can be mathematically described as follows:

$$Forget\ gate: f_t = \sigma(W_f \cdot [h_{t-1}, x_t] + b_f), \qquad (3)$$

$$Input\ gate: i_t = \sigma(W_i \cdot [h_{t-1}, x_t] + b_i), \qquad (4)$$

$$Output\ gate: o_t = \sigma(W_o \cdot [h_{t-1}, x_t] + b_o), \qquad (5)$$

$$Candidate\ state\ values\ calculation: \tilde{C}_t = \tanh(W_C \cdot [h_{t-1}, x_t] + b_C), \qquad (6)$$

$$C_t = f_t \times C_{t-1} + i_t \times \tilde{C}_t, \qquad (7)$$

$$h_t = o_t \cdot \tanh(C_t). \qquad (8)$$

As a type of RNNs specially designed to avoid the exponentially fast decaying factor, the LSTM networks can effectively prevent the gradient vanishing/explosion problem. Due to their ability to learn long term dependencies, LSTMs are particularly suitable for financial prediction problems. Hence, we have the conjecture that extending the linear mean equation to LSTM network will improve the ability of extracting long-term dependencies for duration sequence. To verify this hypothesis, we take the $\Delta t_{i-1}$ and $ln\hat{\mu}_{i-1}$ as the input for the LSTM cell at the time point of i-*th* transaction where $\Delta t_{i-1}$ is the duration of last transaction and $ln\hat{\mu}_{i-1}$ is the logarithmic value of the output of the proposed LSTM-ACD model at time $i-1$. To retain the ability of probabilistic inference, the objective function is still the log likelihood function of $\Delta t_i = \mu_i \varepsilon_i$ which follows an exponential distribution. The log likelihood function can be mathematically described as follows:

$$l = \sum ln \frac{1}{\hat{\mu}_i} exp\left(-\frac{1}{\hat{\mu}_i}\Delta t_i\right) \tag{9}$$

$$ln\hat{\mu}_i = \varphi(\Delta t_{i-1}, ln\hat{\mu}_{i-1}, h_{i-1}) \tag{10}$$

where $\varphi$ represents a mapping from $\Delta t_{i-1}, ln\hat{\mu}_{i-1}, h_{i-1}$ to $ln\hat{\mu}_i$ by a LSTM cell.

**2.2.2 Visualization and promotion by attention mechanism**

Attention mechanism was firstly proposed to improve the image processing accuracy by mimicking the perceptual system of human beings [19]. In the work of [20], attention mechanism was introduced to extend the basic encoder-decoder architecture and enhance the interpretability on the task of machine translation. Unlike the sequence-to-sequence modeling in sentence translation, the problem we focus on in this paper is to predict the financial duration one-step ahead. The attention weights which help automatically search for import hidden states of the sequence-to-one LSTM architecture can be calculated by the following formulas:

$$e_{i-k} = v_\alpha^T tanh(w_\alpha h_{i-k}) \tag{11}$$

$$\alpha_{i-k} = \frac{exp(e_{i-k})}{\sum_{k=1}^{T} exp(e_{i-k})} \tag{12}$$

where $h_{i-k}$ represents the hidden state lagged $k$ time steps and $\alpha_{i-k}$ represents the attention weight of $h_{i-k}$. The $w_\alpha$ and $v_\alpha$ are parameter matrices in the attention mechanism. By allocating different attention weights for different hidden states, a new vector $c_i$ is produced as the input of a feedforward network $f$ for predicting the target variable $y_i$.

$$c_i = \sum_{k=1}^{T} \alpha_k h_{i-k} \qquad (13)$$

$$y_i = f(c_i) \qquad (14)$$

In this study, the attention layer is integrated with LSTM to characterize the dynamics of $\ln \mu_i$ in the above-mentioned mean equation of ACD model. The proposed Attention-LSTM-ACD model can be described by the following equations:

$$h'_{i-k} = LSTM(h'_{i-k-1}, s'_{i-k-1}, \Delta t_{i-k-1}, \ln \hat{\mu}_{i-k-1}) \qquad (15)$$

$$c'_i = \sum_{k=1}^{T'} \alpha'_k h'_{i-k} \qquad (16)$$

$$\ln \hat{\mu}_i = f'(c'_i) \qquad (17)$$

where $s'_{i-k-1}$ represents the cell state of LSTM lagged $k+1$ time steps. Figure 2 shows the Attention-LSTM-ACD model more detailly.

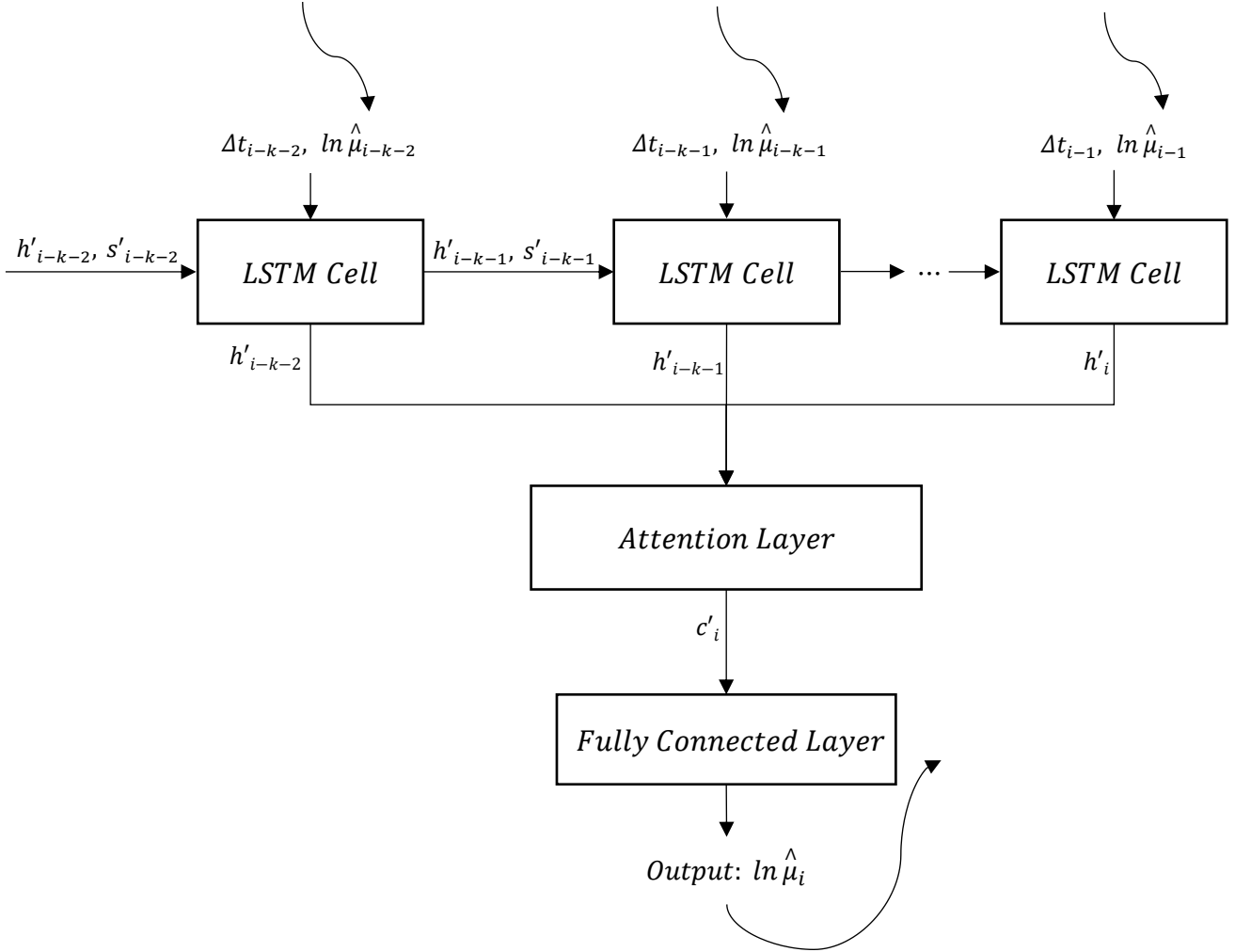

**Fig 2.** Architecture of Attention-LSTM-ACD model

# 3 Experiment

## 3.1 Data description

### 3.1.1 Data source

The Shenzhen Stock Exchange 100 Index (SZSE 100) is the first index that designed for reflecting the multiple-level market conditions of Chinese stock market. The constituent stocks of SZSE 100 represent the core high-quality assets in the Shenzhen A-share market, with strong growth, low valuation and high investment value. In this paper, we collect duration data of the first 100,000 transactions which has excluded the transactions during pre-market opening session, for each constituent stock from SZSE 100. The readers can acquire the data from the Transend DataBase System of Wind Information Co., Ltd (https://www.wind.com.cn/). Since the stock TIANJIN ZHONGHUAN SEMICONDUCTOR CO., LTD., which is coding in 002129.SZ, has no transactions during 2017, we totally have 99 stocks listed in SZSE COMP on December 31st, 2016 as our research dataset, which sums to 9900, 000 transactions.

### 3.1.2. Data Characteristics

As the box plots in Figure 3 demonstrate, transaction durations of each constituent stock from SZSE 100 Index reveals a very long tail compared with the inter-quartile range. The large amount of data locates in the tail means the existence of liquidity risk.

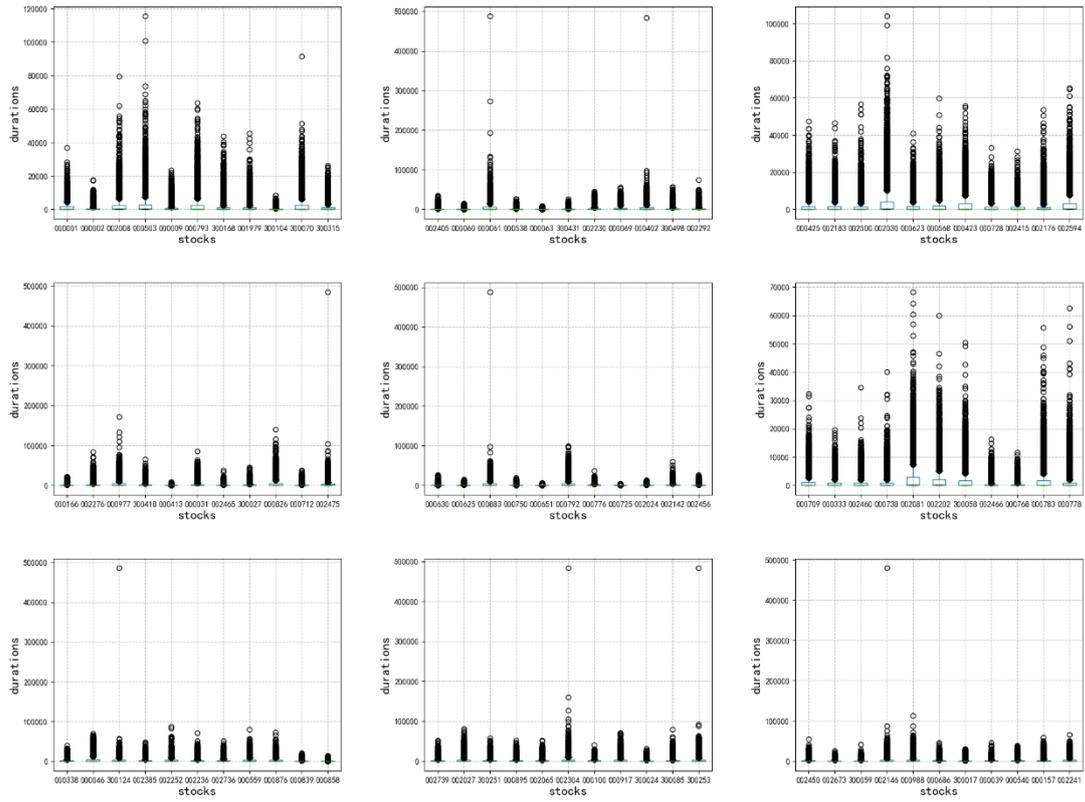

**Fig 3.** Box plots of durations of 99 stocks from SZSE 100 (minimum time unit: millisecond), the 99 stocks are listed in the x-axis and the y-axis represents the duration dimension.

To further dig the dynamic characteristics of the duration sequence, the averaged coefficients of auto correlation function ($acf$) and partial correlation function ($pacf$) coefficients is plotted. As shown in the following Figure.4, we can see that time series duration data shows a longer memory in that both $acf$ coefficients and $pacf$ coefficients decay very slowly as the lagged term increases. Hence, the higher complexity of the UHF duration data requires a forecasting algorithm with strong fitting ability.

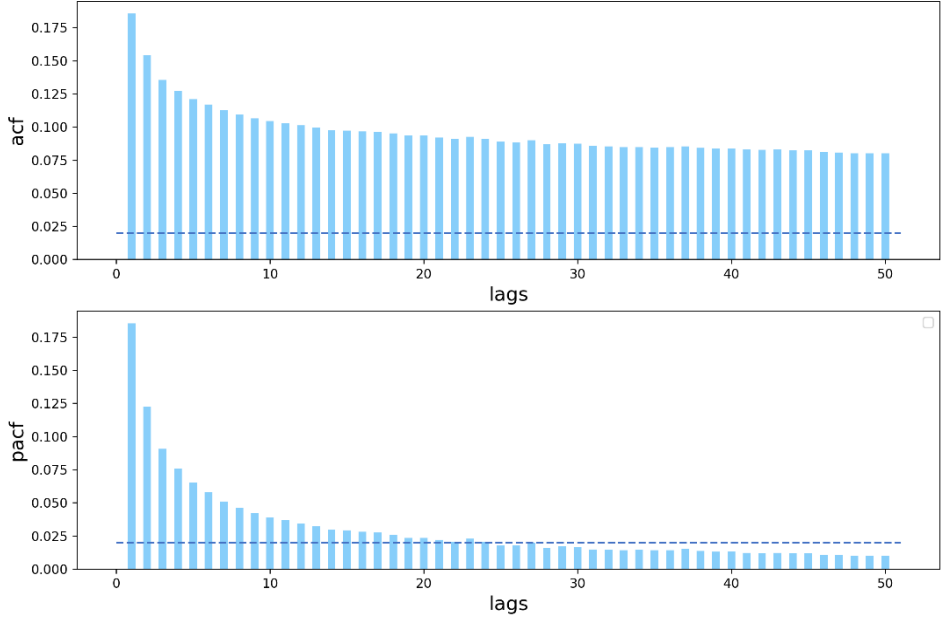

**Fig 4.** Averaged *acf* and *pacf*

## 3.2. Evaluation criteria

### 3.2.1. Mean absolute error (*MAE*)

As one of the most widely used metrics, $MAE$ is used to evaluate the performance of duration prediction more directly and can be calculated by the following formula:

$$MAE = 1/N \sum_{i=1}^{N} |Duration_i^{forecast} - Duration_i^{real}| \qquad (18)$$

A smaller $MAE$ means that we have a more precise forecast of the transaction duration.

### 3.2.2. Performance measure for quantile prediction

To evaluate the forecasting performance of quantile points, we utilize the loss function in quantile regression minimization problems [21]. Let $\{TaR_i : i = 1, \ldots, h\}$ be the prediction quantile points of the confidence level $\alpha$, $x_i$ be the realistic durarion of the $i$-th transaction, and the $I$ represent an indicative function, the performance measure $QL_{\alpha,t}$ can be calculated as follows:

$$QL_{\alpha,t} = \sum_{i=T+1}^{T+h}(x_i - TaR_{i,\alpha})[\alpha - I(x_i - TaR_{i,\alpha})] \qquad (19)$$

### 3.3.3 Experiment models

In Section 2, we have created a new framework for the one-step ahead prediction. The sequence of 50 lagged durations (1 feature, 50 timesteps) is firstly chosen as the input data and we hence construct the LSTM-ACD model and Attention-LSTM-ACD model, which are presented in Section 2.2.1 and Section 2.2.2 respectively. The only difference between the two models is the attention layer. To further utilize the information of transaction by transaction data, one-dimensional duration feature is extended to multi-dimensional feature vector by adding the transaction volume and transaction type information. And then two other models are constructed, named as the Attention-LSTM-ACD(M) model and the LSTM-ACD (M) model. The experiments will be performed with the following five models: the classic ACD model, the LSTM-ACD model, the Attention-LSTM-ACD model, the Attention-LSTM-ACD(M) model and the LSTM-ACD (M) model.

## 3.3. Training

During the training process, configurations are determined with as few exogenous inputs as possible because of the various drawbacks of manual tuning. We adopt fixed hyperparameters including learning rate, number of neurons of each layer, batch size, time steps, etc. for each constituent stock of SZSE 100.

### 3.3.1 Generation of training sets, validation sets and test sets

As mentioned above, the sample used in this study is the 100, 000 durations in 2017 for each stock collected from SZSE 100. We select the last 30% of data as the test set, while the remaining data is divided into training set and validation set according to the ratio of 8:2.

### 3.3.2 Training process

During the experiment, fixed hyperparameter combination is selected for each model based on LSTM-ACD framework. Table 1 lists the hyperparameters used in our experiment. The attention size represents the height of the tensor $w_\alpha$ in formula (11). The initial learning rate is 0.5 and it is reduced by 50% after 1000 training steps. Besides the selection of hyperparameter combination, the remaining parameters of the proposed hybrid models are learned by taking advantage the early-stopping technique to avoid the over-fitting problem. We evaluate model performance on the validation set every 100 training steps and the early stopping patience represents the number of times that there is continuously no improvement in the log likelihood function calculated on the validation set.

**Table 1**: The hyperparameters of each model

|  | Attention-LSTM-ACD (M) & LSTM-ACD (M) | Attention-LSTM-ACD & LSTM-ACD |
|---|---|---|
| Input Layer | 3 features, 50 timesteps | 1 feature, 50 timesteps |
| LSTM layer | 5 hidden neurons | 5 hidden neurons |
| Attention Size | 2 (for model Attention-LSTM-ACD (M)) | 2 (for model Attention-LSTM-ACD) |
| Fully Connected Layer | 2 hidden neurons | 2 hidden neurons |
| Batch Size | 300 | 300 |
| Start Learning Rate | 0.5 | 0.5 |
| Decay Steps | 1000 | 1000 |
| Decay Rate | 50% | 50% |
| Early Stopping Patience | 10 | 10 |

In this paper, the four models based on LSTM-ACD framework are coded in Tensorflow1.0 and the classic ACD method is modelling by the ACDm package based on R language.

### 3.4. Experiment results

#### 3.4.1. Comparison of different models in $MAE$

The out-of-sample forecasting errors of the five types of experiment models are calculated. Table 2 is the average $MAE$ in the test sets when the five models are applied to SZSE 100 Index constituent stocks respectively. The average $MAE$ of the LSTM-ACD (M) model is smaller than the classic ACD model while the remaining three models all perform a bit worse than the classic ACD model. We can also see from the Figure 5 that the LSTM-ACD(M) and LSTM-ACD are both supreme to the classic ACD model on more stocks in the metric of $MAE$. As mentioned above, uniform hyperparameter combination is chosen when applying the hybrid models. If we select different hyperparameters when focusing on different stocks, the performance of these hybrid models will be much better.

In addition, we calculate the $MAE$ of each model for the durations one step lagged by the following formula (20):

$$MAE_{lagged} = 1/N \sum_{i=1}^{N} |Duration_i^{forecast} - Duration_{i-1}^{real}| \qquad (20)$$

The results in the third column in Table 1 show that the average $MAE_{lagged}$ of ACD model is significantly smaller than the average $MAE$. It means the predictions of other four models based on LSTM-ACD framework somewhat convey more meaningful information.

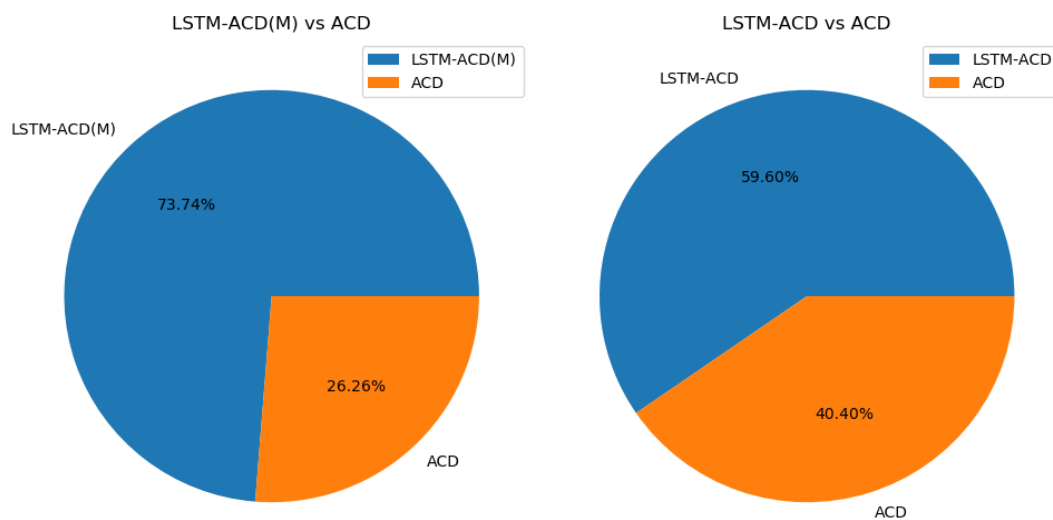

**Fig. 5.** The contrasts of two pairs of models (the left subfigure is the contrast between LSTM-ACD(M) and ACD model, the right subfigure is the contrast between LSTM-ACD and ACD model, the proportion of each slice in a pie chare represents the quantity of stocks on which the corresponding model performs better in $MAE$.)

Table 2: The average $MAE$ on SZSE 100 Index constituent stocks of each model

| Model | Average $MAE$ | Average $MAE$ for durations lagged one step | Difference |
|---|---|---|---|
| Attention-LSTM-ACD (M) | 2.0264 | 1.9762 | 0.0502 |
| LSTM-ACD (M) | **1.7990** | 1.7602 | 0.0388 |
| Attention-LSTM-ACD | 1.9758 | 1.9367 | 0.0390 |
| LSTM-ACD | 1.8964 | 1.8368 | 0.0596 |
| ACD | 1.8641 | 1.6612 | 0.2030 |

### 3.4.2. Comparison of different models in quantile forecasts

Table 3 lists the quantile forecast measure $QL$ of different probability level $\alpha$, for four models. It can be found that the Attention-LSTM-ACD(M) model is the supreme model at all three confidence levels. In terms of the Attention-LSTM-ACD model, it also provides a better quantile forecasting than LSTM-ACD model at all levels. These indicate that the attention layer can improve the accuracy in conditional distribution forecasting.

**Table 3**: Quantile loss for the four models at different $\alpha$ level

| $\alpha = 0.1$ | | | |
|---|---|---|---|
| Attention-LSTM-ACD (M) | LSTM-ACD (M) | Attention-LSTM-ACD | LSTM-ACD |
| **0.7774** | 0.8141 | 0.8084 | 0.8231 |
| $\alpha = 0.05$ | | | |
| Attention-LSTM-ACD (M) | LSTM-ACD (M) | Attention-LSTM-ACD | LSTM-ACD |
| **0.5924** | 0.6578 | 0.6325 | 0.6622 |
| $\alpha = 0.01$ | | | |
| Attention-LSTM-ACD (M) | LSTM-ACD (M) | Attention-LSTM-ACD | LSTM-ACD |
| **0.2988** | 0.3841 | 0.3432 | 0.3892 |

### 3.4.3. Attention weights of different lag orders

This section makes a visualization for the Attention-LSTM-ACD model and Attention-LSTM-ACD (M) model. As can be seen in the following Table 4 and Figure 6, the weights learned by the attention layer in both the two models decrease exponentially with the increase of lag order. That means that the closer transaction has a more important effect on the current duration, which is consistent to our intuition.

**Table 4**: Average weights of the Attention-LSTM-ACD (M) model and the Attention-LSTM-ACD model on SZSE 100 Index constituent stocks

| Attention-LSTM-ACD (M) | | Attention-LSTM-ACD | |
|---|---|---|---|
| Lag order | Weight | Lag order | Weight |
| lag 1 | **0.034797791** | lag 1 | **0.078697926** |
| lag 2 | 0.028296111 | lag 2 | 0.034143126 |
| lag 3 | 0.024825037 | lag 3 | 0.027711418 |
| lag 4 | 0.023690568 | lag 4 | 0.025372255 |
| lag 5 | 0.022575405 | lag 5 | 0.023460835 |
| lag 6 | 0.022012244 | lag 6 | 0.022699354 |
| lag 7 | 0.02148028 | lag 7 | 0.021109585 |
| lag 8 | 0.020997523 | lag 8 | 0.020365109 |
| lag 9 | 0.020678233 | lag 9 | 0.02006378 |
| lag 10 | 0.020491562 | lag 10 | 0.01944402 |
| …… | …… | …… | …… |
| lag 41 | 0.018665664 | lag 41 | 0.017536173 |
| lag 42 | 0.018792729 | lag 42 | 0.017302261 |
| lag 43 | 0.018596823 | lag 43 | 0.017404814 |
| lag 44 | 0.018732648 | lag 44 | 0.017530387 |
| lag 45 | 0.018695344 | lag 45 | 0.017503974 |
| lag 46 | 0.018753901 | lag 46 | 0.017434779 |
| lag 47 | 0.018770904 | lag 47 | 0.017377114 |
| lag 48 | 0.01898069 | lag 48 | 0.017801802 |

| lag 49 | 0.018924108 | lag 49 | 0.017549126 |
| lag 50 | 0.018977175 | lag 50 | 0.01779628 |

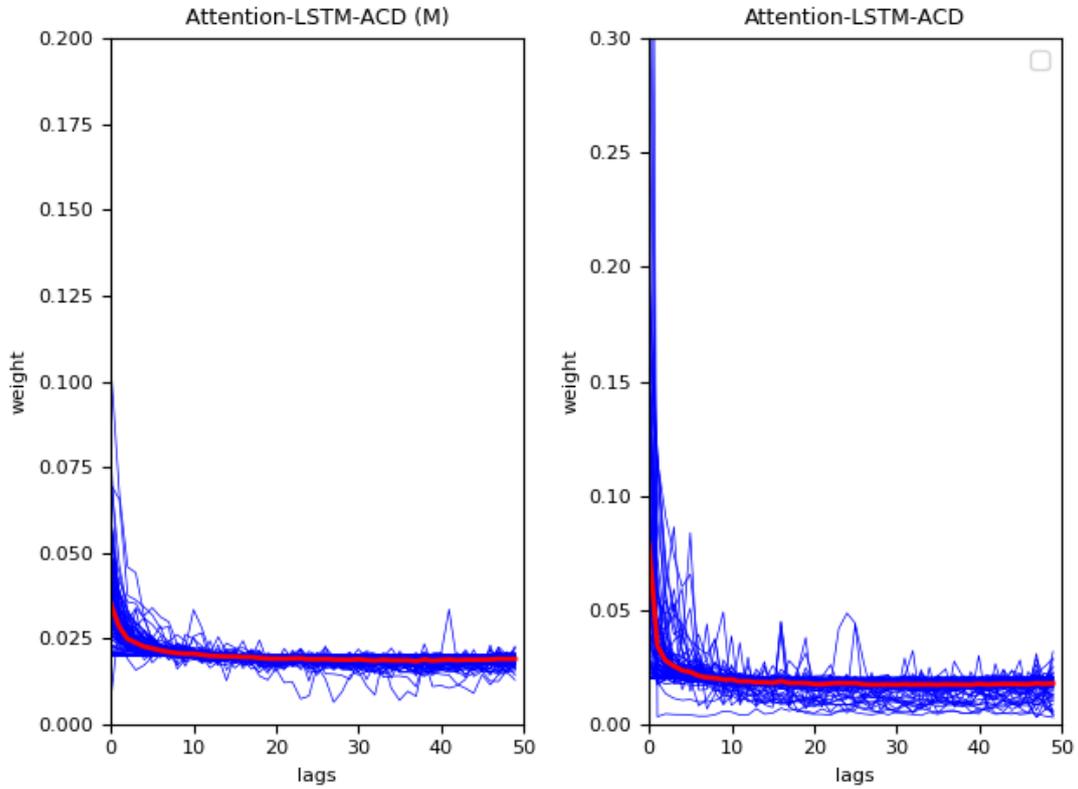

**Fig 6.** Attention weights of the Attention-LSTM-ACD (M) model and Attention-LSTM-ACD model of different lags on SZSE 100 Index constituent stocks (each bule line represents the attention weight sequence of a stock for the corresponding model, each red line represents the average attention weights on SZSE 100 Index constituent stocks for the corresponding model)

## 4 Conclusion and future research

In this paper, we review the studies of transaction duration modelling based on ACD framework and find that these studies can be classified into two categories: a. propose a new non-linear equation form to describe the dynamics of conditional (expected) duration; b. choose a more flexible distribution for the unexpected portion of the duration.

This study constructs a new framework for transaction duration modeling from the perspective of extending the mean equation of ACD model by machine learning methods. Firstly, we build a LSTM-ACD model by combining the LSTM networks with classic ACD model to characterize the complexity of the conditional mean process while retaining the advantage of providing probabilistic output. And then, attention layer is added to construct the Attention-LSTM-ACD model with the ability of unveiling importance of each hidden state in the LSTM networks.

Our proposed new framework is applied to a large-scale dataset. The fixed hyperparameters are chosen for all constituent stocks of SZSE 100 index to reduce the impact of manual tuning and the parameters (and consequently the underlying distributions) are learned via maximize the log-likelihood function. The results show that LSTM-ACD (M) model can present highest accuracy on the task of forecasting on real-world financial datasets among all the presented models. Although Attention-LSTM-ACD model and Attention-LSTM-ACD (M) model could not provide a more accurate performance in $MAE$ metric, the attention layer vividly depicts the importance of different temporal points of the input sequence and outperforms the corresponding LSTM-ACD model and LSTM-ACD (M) model in $QL$ loss metric respectively. In addition, the average $MAE_{lagged}$ of ACD model is significantly smaller than the average $MAE$, which means the predictions of the LSTM-ACD framework models to some extent convey more meaningful information. As a suitable chosen residual distribution does matters, the exponential distribution used in our framework can be extended to more flexible distributions in future research.